\documentclass[runningheads,anonymous]{llncs}
\usepackage{graphicx}

\usepackage[most]{tcolorbox}
\usepackage{enumitem}
\usepackage{hyperref}
\usepackage{array}

\newcommand{\ia}{\textit{i}}
\newcommand{\ib}{\textit{ii}}
\newcommand{\ic}{\textit{iii}}

\begin{document}
\title{A Grassroots Architecture to Supplant Global Digital Platforms by a Global Digital Democracy}

\titlerunning{Grassroots Architecture}
\author{Ehud Shapiro}
\authorrunning{Shapiro}
\institute{Weizmann Institute of Science and London School of Economics}
\maketitle              % typeset the header of the contribution
\begin{abstract}
We present an architectural alternative to global digital platforms termed \emph{grassroots}, designed to serve the social, economic, civic, and political needs of local digital communities, including their federation.  Grassroots platforms may offer local communities an alternative to global digital platforms while operating solely on the smartphones of their members, forsaking any global resources other than the network itself.  Such communities may form digital economies without initial capital or external credit, exercise sovereign democratic governance, and federate, ultimately resulting in the grassroots formation of a global digital democracy.

\end{abstract}
\section{Introduction}\label{section:overview}

The digital realm today is dominated by global platforms:
Centralised systems that have adopted surveillance-capitalism~\cite{zuboff2019age} as their business model, e.g. Facebook; and decentralised systems that implement global cryptocurrencies and contribute to the flow of wealth from the multitudes to the few, e.g. Bitcoin~\cite{bitcoin}. Global digital platforms exacerbate inequality and undermine the fabric of human society by depleting the social, economic, civic, and political capital of local communities and consequently of whole countries, worldwide.

Here we present an architectural alternative to global digital platforms termed \emph{grassroots}~\cite{shapiro2023grassroots,shapiro2023grassrootsBA}.  It is designed to serve the social, economic, civic, and political needs of local digital communities, as well as provide for their federation.  The architecture support grassroots platforms that may offer local alternatives to global digital platforms while operating solely on the smartphones of their members, forsaking any global resources other than the network itself. The grassroots architecture presented here is a major revision and expansion of an earlier proposal with a similar goal~\cite{shapiro2022foundations}. 

The grassroots architecture incorporates a grassroots protocol stack and employs the blocklace---a generalisation of the blockchain that functions as a universal, Byzantine fault-tolerant, Conflict-free Replicated Data Type (CRDT)~\cite{almeida2024blocklace}. The grassroots protocol stack provides for dissemination~\cite{shapiro2023grassroots}, equivocation exclusion~\cite{keidar2023cordial,lewispye2023flash}, ordering consensus~\cite{keidar2023cordial,lewis2024goodfellas}, State Machine Replication/digital social contracts~\cite{cardelli2020digital}, democratic governance~\cite{shapiro2022foundations}, and federation~\cite{halpern2024federated,halpern2024grassroots}.
Envisioned grassroots platforms include grassroots social networks~\cite{shapiro2023gsn}, grassroots currencies~\cite{shapiro2024gc}, and a grassroots digital economy comprising people as well as legal persons (banks, corporations, cooperatives, communities, municipalities, and states) employing off-chain and on-chain governance.  Together, the grassroots protocol stack and platforms aim to provide scalable foundations for grassroots digital communities that emerge locally and federate globally~\cite{halpern2024grassroots}.  Such communities may form digital economies without initial capital or external credit, exercise sovereign democratic governance, and federate, ultimately resulting in the grassroots formation of a global digital democracy~\cite{shapiro2022foundations,halpern2024federated,halpern2024grassroots}.
\begin{figure}[ht]
  \begin{center}
   \includegraphics[width=12cm]{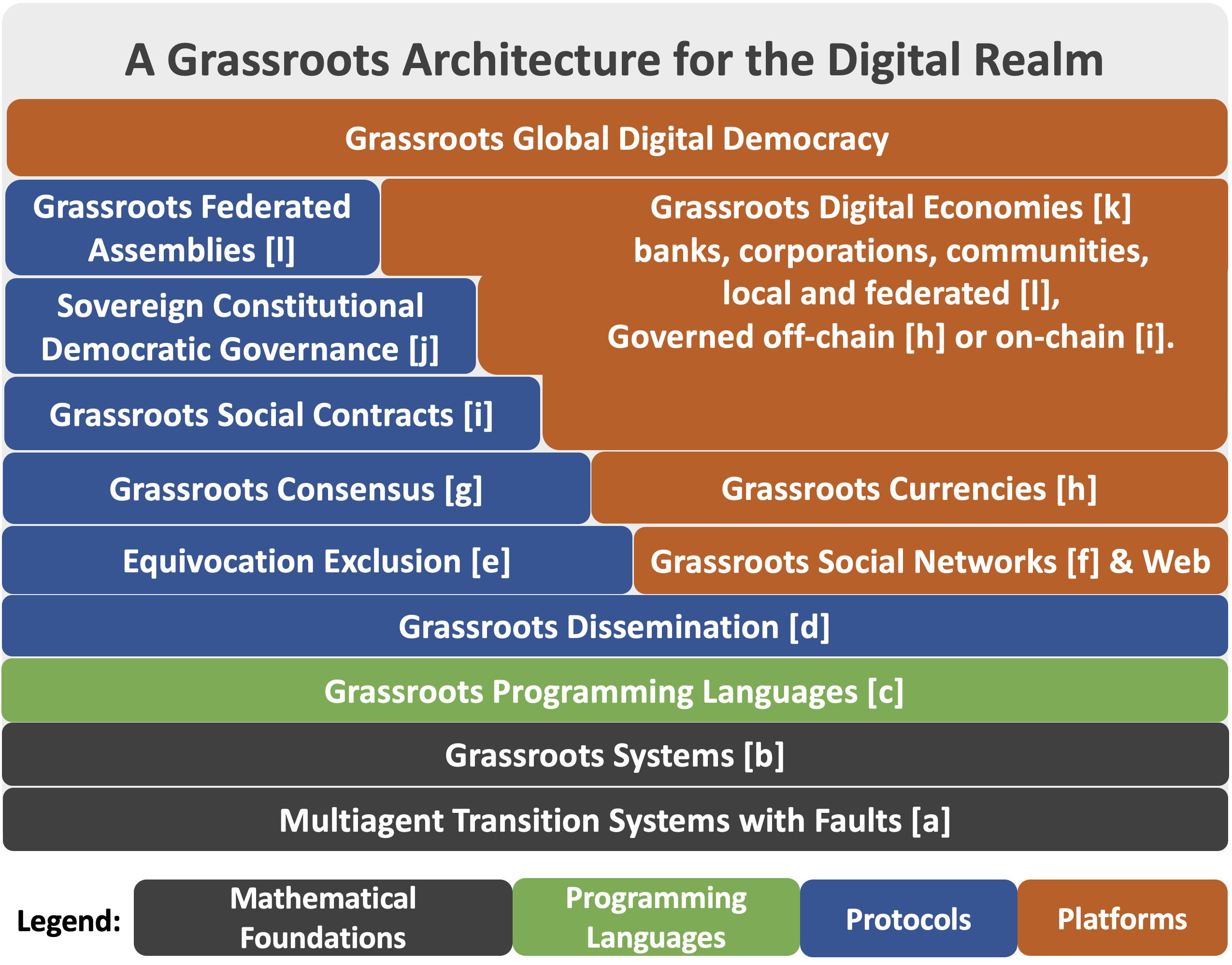}
  \end{center}
\caption{A Grassroots Architecture for the Digital Realm.   Bold letters map to references as follows: 
  \textbf{[a]}$\mapsto$\cite{shapiro2021multiagent},
  \textbf{[b]}$\mapsto$\cite{shapiro2023grassroots,shapiro2023grassrootsBA},
  \textbf{[c]}$\mapsto$\cite{shapiro2024gpl},
  \textbf{[d]}$\mapsto$\cite{shapiro2023grassroots,shapiro2023grassrootsBA,shapiro2023gsn,almeida2024blocklace},
  \textbf{[e]}$\mapsto$\cite{lewispye2023flash,lewis2023grassroots,almeida2024blocklace},
  \textbf{[f]}$\mapsto$\cite{shapiro2023gsn}
  \textbf{[g]}$\mapsto$\cite{keidar2023cordial},
  \textbf{[h]}$\mapsto$\cite{shapiro2024gc,lewis2023grassroots},
  \textbf{[i]}$\mapsto$\cite{cardelli2020digital},  \textbf{[j]}$\mapsto$\cite{abramowitz2021democratic,abramowitz2021beginning,bulteau2021aggregation,elkind2022complexity,elkind2021united,meir2020sybil,shahaf2019sybil,shapiro2018incorporating,shapiro2018incorporating},
  \textbf{[k]}$\mapsto$\cite{shapiro2024gc},
\textbf{[l]}$\mapsto$\cite{halpern2024federated,halpern2024grassroots}.
  }
\label{figure:architecture}
\end{figure}
In this paper we review the core concepts of the architecture---digital sovereignty, grassroots systems (Figure \ref{figure:architecture}, top black box), mention the desiderata on grassroots programming languages (ibid., green box), explain the blocklace and the grassroots protocol stack (ibid., blue boxes), and only reference the grassroots platforms (ibid., brown boxes).

\section{Architecture Core Concepts}\label{section:foundations}

The grassroots architecture is designed to be implemented on people's smartphones, with server-farms up in the clouds only playing an auxiliary role, as needed.  In that regard it is worth noting that today's smartphones have thousands of times more of the computing power, memory capacity and networking speed of the Unix workstations that were the workhorses of the Internet and the web at their inception.  The reasons that today's role of smartphones on the Internet is only to access the platforms that operate on it, not to run them, are historical, economic, political, and architectural, not their performance.  The grassroots architecture aims to remove these barriers through the notion of grassroots platforms.

The grassroots architecture incorporates three concepts at its core:
(\ia) Digital sovereignty, the notion that people and communities should have control over their digital lives commensurate with the control they have over their physical lives; (\ib)
Grassroots systems, mathematically capturing the inherent nature of the architecture; and (\ic) The blocklace, serving as its basic and universal data structure.  We review each in turn.

\subsection{Digital sovereignty}\label{subsection:sovereinty}
The notion that people should have sovereignty over their digital personae has been pursued most avidly under the moniker \emph{self-sovereign digital identity}~\cite{allen2016path,RN47,RN592}.   For digital communities, the notion is manifest most clearly in  \emph{freedom of digital assembly}, the ability of people to assemble digitally using their  networked personal computing devices (e.g. smartphones), and not at the behest of a third party [\href{https://images.app.goo.gl/bRfM1xhfhPwzuhyB6}{See KAL's cartoon}].  

Digital sovereignty in general, and digital freedom of assembly in particular, are a precondition for attaining equality in democratic digital governance:  Without them, a third party with control over a digital community might collude with some members of the community to their advantage, hampering political equality.  Grassroots systems, discussed next, offer the computational foundation needed to provide individuals and communities with digital sovereignty in general and freedom of assembly in particular.

\subsection{Grassroots Systems}\label{subsection:systems}
Informally, a \emph{grassroots system} is a distributed system that can have multiple instances, independent of each other and of any global resources,  that may interoperate once interconnected~\cite{shapiro2023grassroots,shapiro2023grassrootsBA}.
Global platforms are not grassroots:  They are designed for global dominance and be one of a kind:  One Facebook, one Bitcoin.  Global platforms that are similar, e.g. Bitcoin, Bitcoin Cash and Bitcoin SV, are designed to compete with each other rather than to interoperate. 

BitTorrent is an example of a grassroots system, albeit server-based.  Grassroots platforms aim to be serverless, and operate solely on the networked smartphones of their members.
An example grassroots system would be a serverless smartphone-based  social network supporting multiple independently-budding communities that can merge when a member of one community becomes also a member of another~\cite{shapiro2023gsn}.

The notion of grassroots systems and grassroots implementations was formalised using the mathematical framework of \emph{multiagent transition systems}~\cite{shapiro2021multiagent}.  The specification included an abstract grassroots dissemination protocol; described and prove an implementation of grassroots dissemination for the model of asynchrony; extend the implementation to mobile (address-changing) devices that communicate via an unreliable network (e.g. smartphones using UDP).  Grassroots platforms that were specified formally, but not yet implemented, include grassroots social networks~\cite{shapiro2023gsn} and grassroots currencies~\cite{shapiro2024gc}.
The mathematical construction employs distributed multiagent transition systems to define the notions of  grassroots protocols and grassroots implementations,  to specify grassroots dissemination protocols and their implementation, and to prove their correctness.  The implementations use the blocklace, described next.

\subsection{Grassroots Programming Languages}\label{subsection:systems}

A programming language for grassroots systems~\cite{shapiro2024gpl} should serve two goals:  First, to ease the implementation of grassroots protocols, grassroots applications and grassroots platforms, by providing an abstraction of the underlying machinery they necessitate.  Second, to ease the proof that the resulting system is indeed grassroots.   To do so, a grassroots programming language must gave at least four characteristics:
\begin{enumerate}
    \item Be a concurrent programming language, supporting the dynamic creation of processes, their communication and synchronisation.
    \item Its processes must be independent of any global/shared/third party resource for their execution.
    \item Processes can reconfigure their communication network dynamically.  For this,  communication channels should be `first class objects', so that one process can send another a message that contains a communication channel with a third process, and the recipient can then communicate with that third process.
    \item Programs should also be `first class objects', so that one process can send a message that contains a program to another, which can then execute the program upon receipt.
\end{enumerate}
In addition, ideally a grassroots programming language should support metaprogramming~\cite{sterling1994art,shapiro1989family,safra1988meta,lichtenstein1988concurrent}, to ease the implementation of a grassroots operating system and program development environment for the grassroots programming language  within the language.  We are presently exploring addressing this challenge with a single-reader/single-writer subset of Concurrent Prolog~\cite{shapiro2024gpl,shapiro1989family,shapiro1987concurrent}.

\subsection{The Blocklace}\label{subsection:blocklace}
\begin{figure}[ht]
  \begin{center}
   \includegraphics[width=9cm]{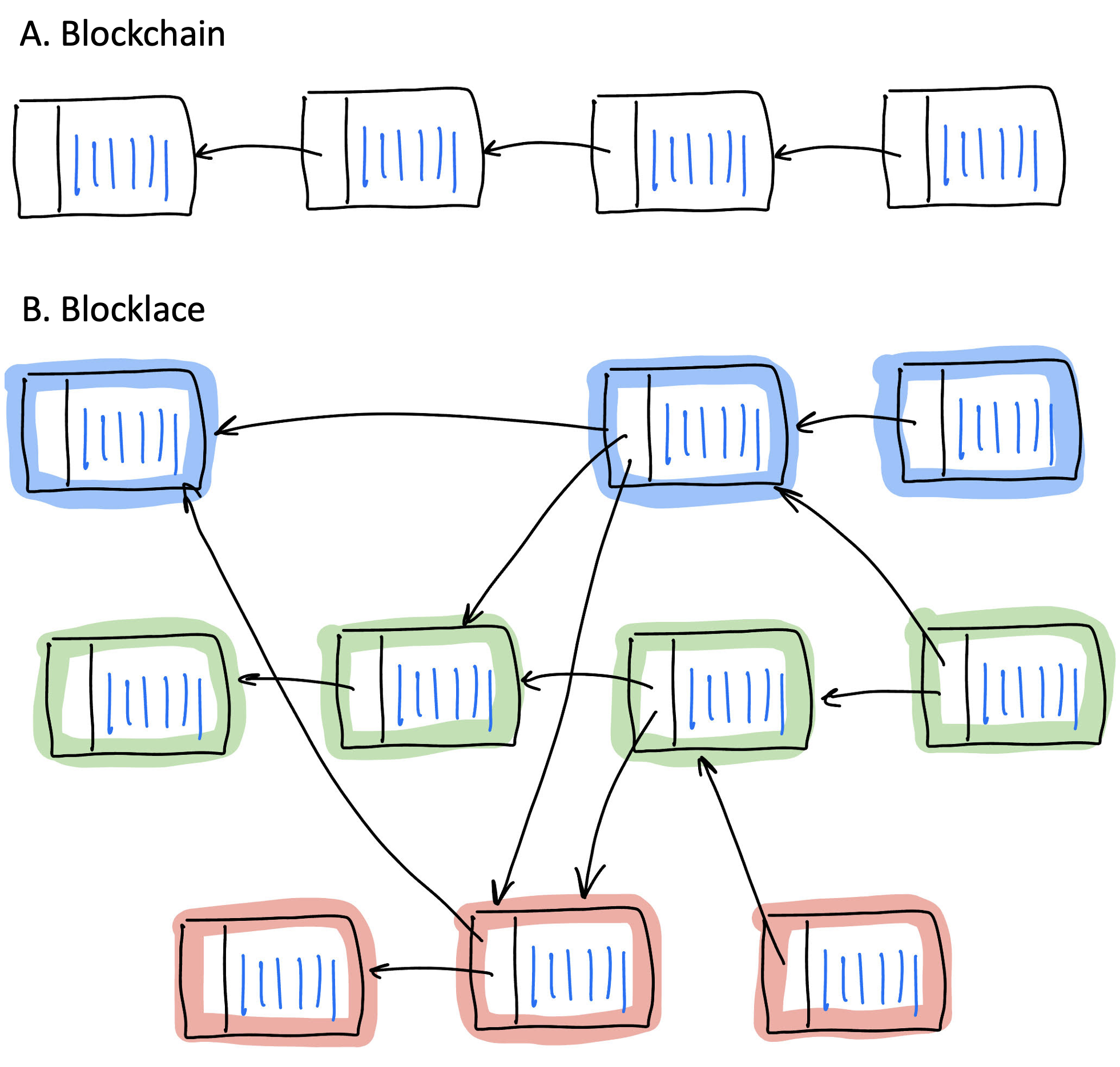}
  \end{center}
\caption{\textbf{A. Blockchain.}  The first block on the left is the \emph{genesis block}.  The blockchain is (\ia) \textit{tamper-proof}, as a payload or pointer cannot be changed without being detected; (\ib) \textit{non-repudiable}, as a signed block or payload cannot be repudiated by the block's signatory, and  (\ic) \textit{totally-ordered}.\newline
\textbf{B. Blocklace.}   Agents produce their own (colour-coded) signed genesis blocks and interlinked \emph{virtual blockchains} (e.g., the red blocks by themselves do not form a blockchain). 
The blocklace is similarly  (\ia) tamper-proof and (\ib) non-repudiable, but is (\ic) \textit{partially-ordered}.
  }
\label{figure:blockchain-blocklace}
\end{figure}

The \emph{blockchain} (Figure \ref{figure:blockchain-blocklace}.A) is a linear data structure in which each block except the first has a hash pointer to its predecessor.  It is constructed by agents competing to add the next block to the blockchain.
Blockchain consensus protocols such as Nakamoto
Consensus~\cite{bitcoin} resolve the conflicts, or `forks', that arise
when two or more agents add their own blocks to the same blockchain, in order to produce the consensual linear blockchain. 

The \emph{blocklace} (Figure \ref{figure:blockchain-blocklace}.B, aka block DAG) is a partially-ordered generalisation of the blockchain in which
each block has any finite number of signed hash pointers to preceding blocks. It can
be constructed cooperatively, by agents adding blocks to the blocklace
and informing each other of new blocks. 

The description above suggests that while the blockchain is fundamentally a
conflict-based datatype, the blocklace is not. 
Conflict-free Replicated Data Types (CRDTs) are designed for convergent construction without global coordination or consensus. The blocklace datatype, with the sole operation of adding a single block, can be shown to be a CRDT~\cite{DBLP:conf/sss/ShapiroPBZ11}. Allowing arbitrary values as payload, the blocklace can also be seen as a universal Byzantine fault-tolerant implementation for arbitrary CRDTs~\cite{DBLP:journals/corr/abs-2012-00472,DBLP:conf/eurosys/Kleppmann22,DBLP:conf/sicherheit/JacobBH22}.

Previous applications of the blocklace datatype include grassroots dissemination~\cite{shapiro2023grassroots,shapiro2023grassrootsBA}, grassroots social networks~\cite{shapiro2023gsn},  a payment system~\cite{lewispye2023flash}, an implementation of grassroots currencies~\cite{lewis2023grassroots},  a family of Byzantine Atomic Broadcast consensus protocols~\cite{keidar2023cordial}, and multiple improvements and extensions to the Nakamoto Consensus protocol (under the name block DAG)~\cite{lewenberg2015inclusive,sompolinsky2021phantom,sompolinsky2016spectre,sompolinsky2022dag}.

\section{The Grassroots Protocol Stack}\label{section:protocol-stack}

A consensus protocol requires achieving \emph{dissemination} (letting everyone know of every block), \emph{equivocation exclusion} (preventing double-spending in digital payment systems and cryptocurrencies) and \emph{ordering}  (for State Machine Replication in general and smart contracts in particular).

Often the three tasks are addressed by a single protocol, e.g. Nakamoto Consensus~\cite{bitcoin}.  The grassroots protocol stack addresses each task by a distinct protocol (See protocol stack in blue in Figure \ref{figure:architecture}). 
Each protocol layer supports both the layer above it in the stack, as well as applications that can be implemented more efficiently, or even only by, using that layer's protocol rather than the entire protocol stack.

The dissemination layer supports grassroots dissemination, which in turn can support grassroots social networks (brown box, ibid.), and is more general than the all-to-all dissemination required by consensus.  The equivocation exclusion layer support both supermajority-based equivocation exclusion, required for payment systems~\cite{lewispye2023flash} and for consensus in general, and also the weaker notion of leader-based equivocation exclusion, sufficient to realise grassroots currencies~\cite{lewis2023grassroots} (another brown box, ibid.).
Finally, the ordering consensus layer supports State Machine replication/digital social contracts~\cite{cardelli2020digital}, which in turn support on-chain democratic governance of digital communities and their federations (blue boxes ibid.).
All these notions are elaborated below.

\subsection{Grassroots Dissemination}\label{subsection:dissemination}
The basic tenet of \emph{grassroots dissemination}~\cite{shapiro2023grassroots} is that an agent that needs a block can obtain it from a friend that has it. This abstract notion is realised by employing the blocklace as follows:
\begin{enumerate}[leftmargin=*]
    \item \textbf{Disclosure}: Inform your friends which blocks you know and which you need.
    \item \textbf{Cordiality}: Send to your friends blocks you know and think they need.
\end{enumerate}

Disclosure is realised by each correct agent $p$ incorporating in every new $p$-block hash pointers to the current tips of $p$'s local blocklace.  Thus a new $p$-block serves as a multichannel ack/nack:  It ack's blocks known to $p$ so far, and by the same token nack's blocks not yet known to $p$. Cordiality is realised by each correct agent $p$ sending to any friend $q$ any block in the set-difference between  the local blocklace of $p$ and the set of blocks \emph{referred} to (by a chain of one or more pointers)  by the most-recent $q$-block received by $p$.  
Communication is only among friends, as specified  by the social graph,  encoded by special blocks in the blocklace.  Different grassroots platforms employ different types of social graphs, as shown in Table \ref{table:social-graphs}.

\begin{table}[!ht]
%  \begin{center}
  \small 
 \begin{tabular}{ | m{7em} | m{9em}| m{10em} | m{10.5em} | } 
    \hline
    \textbf{Protocol/\newline Social Graph} & \textbf{Grassroots Twitter-Like}~\cite{shapiro2023gsn}
 & \textbf{Grassroots WhatsApp-Like}~\cite{shapiro2023gsn}  & \textbf{Grassroots} \newline\textbf{Cryptocurrencies}~\cite{lewis2023grassroots}
\\
     \hline
    \hline
    \textbf{Graph type} & Directed  & Hypergraph 
                                                    & Undirected  \\
     \hline
    \textbf{Edge meaning} & Source agent follows destination agent
     & A group of agents  & Mutual credit line\\
 \hline
    \textbf{Friendship \newline condition} & Mutual following & Membership in same group & Mutual credit offers
\\
    \hline

    \textbf{Graph update}  & Agent follows/unfollows another agent & Agent founds group;  invites/removes agents & Agents make/break friendships
\\
 \hline 
  \textbf{Agent needs}  &  Feeds of agents it follows &  Feeds of groups it is a member of
   &  Payments to it/in its coin; approvals of payments by/to it.\\
    \hline 
  \textbf{Data \newline  Structure }  &  Blocklace &  Blocklace   &  Blocklace
\\
 \hline   
  \end{tabular}
%  \end{center}
   \caption{Social Graphs for Grassroots Social Networks and Grassroots Currencies}
 \label{table:social-graphs}
\end{table}
The grassroots Twitter-like protocol  employs a directed graph.  A directed edge $p \rightarrow q$ means that $p$ \emph{follows} $q$. Any agent can create and remove outgoing edges at will.  We say that $p$ and $q$ are \emph{friends} if the social graph has edges $p \leftrightarrow q$.
The grassroots WhatsApp-like protocol employs a hypergraph (in which an edge may connect any number of vertices).  A hyperedge connecting a set of agents means that the agents are members in a \emph{group} represented by the hyperedge.  
We say that $p$ and $q$ are \emph{friends} if they are members of the same hyperedge of the social graph.
Any agent $p$ can create a group with $p$ as its sole member. A group creator can invite other agents to become members and remove members at will.  An invited agent may join the group and leave it at will.
In grassroots currencies, friendship is manifest through a mutual line of credit, in which each agent holds grassroots coins  (digital IOUs) issued by the other.

\subsection{Equivocation Exclusion}\label{subsection:eqivocation}
\emph{Double-spending} occurs when an agent pays the same digital coin to two different agents.
Preventing double-spending is the major task of digital payment systems, including implementations of cryptocurrencies.
\emph{Equivocation} is the means to realise double-spending in a blockchain/blocklace-based system, by an agent creating two blocks that do not refer to each other.   If equivocating blocks contain a double-spending,  the recipient of each block may be misled to think that they have received a valid payment of the coin.
Hence, correct agents does not equivocate, namely a new block by an agent refers to their previous block (except for the first such block).

Equivocation exclusion is typically realised by the Reliable Broadcast protocol~\cite{bracha1987asynchronous,guerraoui2019consensus,auvolat2020money,collins2020online}.  The blocklace allows more efficient equivocation exclusion, based on \emph{supermajority approval}.
Given a set of $n$ agents, at most $f$ of which are \emph{faulty} and the rest are \emph{correct}, a \emph{fault-resilient supermajority}, or \emph{supermajority} for short, is any fraction of agents greater than $\frac{n+f}{2}$.  For example,  if $f=0$ then a simple majority is a supermajority, and if $f<\frac{1}{3}$ then  $\frac{2}{3}$ is a supermajority. Its two key properties are: (\ia) A supermajority includes a majority of the correct agents. (\ib)  Two supermajorities must have a correct agent in common.

Supermajorities support equivocation exclusion as follows. An agent $p$ \emph{approves} a block $b$ if there is a $p$-block that refers to $b$ but not to any block equivocating with $b$.  It can be shown that an agent cannot approve two equivocating blocks without equivocating itself, which implies that two equivocating blocks cannot both receive supermajority approval (assume they do: then the two supermajorities must have a correct agent in common, which must equivocate to approve the two equivocating blocks; a contradiction).

Blocklace supermajority-based equivocation exclusion was employed by the Cordial Miners consensus protocol~\cite{keidar2023cordial} and the Flash payment system~\cite{lewispye2023flash}. The Grassroots Flash payment system~\cite{lewis2023grassroots} employs the blocklace but with leader-based equivocation exclusion, where the issuer of a coin resolves equivocations involving payments in their coin.

\subsection{Ordering Consensus}\label{subsection:consensus}
DAG-based consensus protocol are all the rage~\cite{keidar2021need,giridharan2022bullshark,keidar2023cordial,babel2023mysticeti}, as they offer $O(n)$ throughput---allowing all agents to produce blocks simultaneously on every round---compared to $O(1)$ throughput of blockchain-based or leader-based consensus protocols.  Among DAG-based protocols, blocklace-based protocols such as Cordial Miners~\cite{keidar2023cordial} offer improved latency, as they forsake Reliable Broadcast in favour of Cordial Dissemination and supermajority-based equivocation exclusion, discussed above.  The protocol orders the partially-ordered blocklace by a topological sort of the equivocation-free DAG, ensuring that all agents resolve any indeterminacy in the partial order uniformly and thus produce the same total order.

Further improving latency and complexity using the blocklace while maintaining high throughout is possible~\cite{lewis2024goodfellas}.  But for local communities employing social contracts for democratic governance such high throughput is not needed and its overhead is untenable.  Hence low-throughput blocklace-based consensus protocols, suitable for operating on the smartphones of grassroots communities, are being developed~\cite{lewis2024laidback}.

\subsection{Digital Social Contracts}\label{subsection:social-contracts}
Ordering consensus enables State Machine Replication.  In case of a blockchain  supporting a cryptocurrency, such as Ethereum~\cite{buterin2014next}, blockchain consensus further enables \emph{smart contracts}, which are replicated state programs that may include cryptocurrency transactions~\cite{buterin2014next,de2021smart,wang2018overview}.  Smart contracts may specify the accounts that can participate, or activate, the contract, and are executed by third-parties external to the contract---the agents, or `miners',  that execute the blockchain consensus protocol. These agents are remunerated for their efforts via cryptocurrency payments referred to as \emph{gas}.

Adhering to digital sovereignty requires achieving the functionality of a smart contract without paying the nominal and conceptual fee to the third parties. This is achieved by Proudhon's 1851 notion of the \emph{social contract:} a voluntary agreement among free individuals that does not require one individual to surrender sovereignty to another ~\cite{proudhon2004general}. 
We define  a \emph{digital social contract} to be Proudhon's notion of the social contract transported to the digital realm -- \emph{a voluntary agreement among free individuals, specified, undertaken, and fulfilled in the digital realm}~\cite{cardelli2020digital}. We identify a sub-class of digital social contracts---\emph{consensus-based digital social contracts}---as the sovereign counterpart of smart contracts. In a smart contract, the execution of the underlying ordering consensus protocol, which enables the State-Machine Replication needed for smart-contract execution, is relegated to anonymous third-party miners of the underlying cryptocurrency. In contrast, a consensus-based digital social contract is executed by its parties,  and hence does not owe `gas' or depend on anonymous third-party miners for its existence. Thus, the parties to a consensus-based digital social contract are jointly and equally its sovereigns.

The quintessential example of a smart contract is a DAO~\cite{ethereum:dao}. In the context of cryptocurrencies, the governance of DAOs is typically plutocratic, as parties to the contract participate and vote with their digital wallets. In contrast, a digital social contract among vetted participants that specifies a DAO can be democratic. Furthermore, such a digital social contract  is truly-Autonomous (the `A' in DAO), as opposed to the standard DAO, which depends upon, and has to remunerate, anonymous third-party miners for its execution.  Thus,  digital social contracts can be egalitarian and a truly-Autonomous counterpart of DAOs.  Many of the criticisms of blockchain-based democracy~\cite{tromer2018benaloh} evaporate when the citizens are also the `miners', as there is no divergence of interests between the miners and the voters.  Combined with smartphone-based consensus, digital social contracts may be the engine of on-chain democratic governance of grassroots communities, discussed next.

\subsection{Democratic Governance of Digital Communities}\label{subsection:democracy}
Democracy is not an end in itself, but rather a means for people to have equal standing in all aspects of the political decision-making process.  Achieving equality in 
the governance of digital communities is a multifaceted challenge:
\begin{enumerate}
  \item \textbf{Equality in voting:} Sybils, namely fake and duplicated digital identities, are a prime threat to equality~\cite{alvisi2013sok,douceur2002sybil,seuken2014sybil,siddarth2020watches,borge2017proof}, as single sybil they may subvert a democratic decision. Reality-aware social choice~\cite{shapiro2018incorporating}, augmented so that the status quo can be changed only with supermajority support, was shown to be resilient to both bounded sybil penetration and to partial participation of the correct agents~\cite{meir2020sybil,meir2024safe}. 
    \item \textbf{Equality in proposing:} Incorporating a metric spaces in reality-aware social choice,  with the assumption that voter preferences are single-peaked, can makes voting and proposing the same act--- stating the most-preferred point in the metric space. This provides a unified approach  to making participants equal as both voters and proposers in single-winner elections, committee elections,  budgeting and legislation~\cite{bulteau2021aggregation}.
    \item \textbf{Equality in deliberation and coalition formation:} Adding to reality-aware social choice
    with metric spaces the dynamics of coalition formation, in which participants may form coalitions around proposals, offer compromise proposals, and switch from one coalition to another, allows participants to cooperatively search a metric space for a proposal most-supported over the status quo~\cite{elkind2021united,elkind2022complexity}. 
    \item \textbf{Equality in constitution formation and amendment:} Inspired by May's Theorem~\cite{may1952set},  an egalitarian  Condorcet-consistent initial constitutional amendment rule was derived from first principles~\cite{abramowitz2021beginning}.
    \item \textbf{Equality in community forking:} 
     Cryptocurrencies employ informal governance that often results in \emph{hard forks}~\cite{webb2018fork}, in which a protocol is abandoned by some or all of the participants. \emph{On-chain governance}~\cite{reijers2018now} aims to avoid hard forks by providing a decision process in which the cryptocurrency's protocol can be amended `from within' according to the rules of the protocol itself.  Reality-aware social choice turns the chaotic and often plutocratic process of digital community forking into a democratic process~\cite{abramowitz2021democratic}.
\end{enumerate}
\subsection{Grassroots Federated  Assemblies for Global Democratic Governance}\label{subsection:federation}
The democratic governance of large-scale digital communities is an open problem.  Key challenges include:
\begin{enumerate}
\item \textbf{Sybils:} The penetration of fake and duplicate digital identities.  Facebook, for example, removes around 1Bn fake accounts every quarter [\href{https://www.statista.com/statistics/1013474/facebook-fake-account-removal-quarter/}{link}].
\item \textbf{Large-scale online voting:} Many leading experts consider large-scale online voting to be untenable~\cite{park2021going}, [\href{https://www.aaas.org/epi-center/internet-online-voting}{link}].
\end{enumerate}
An effort to address this problem employs grassroots federation~\cite{halpern2024federated,halpern2024grassroots}.  A federation may include people and/or child federations forms voluntarily and is governed by a small (say 100-member) assembly chosen by sortition from its individual members and the members of the assemblies of its child federations.  The assembly of a federation is sovereign to decide on the admission and removal of its members, as well as on the application to join higher-level federations.  The strata to be represented by the assembly of a federation are determined by the evolving structure of the federation and sizes of its member assemblies, which may intersect and therefore have to be weighted.

This approach addresses sybils by having the federation built from small local communities in which members know each other to be genuine, and from communities that federate only if they trust each other to be genuine,  for example by having sufficient intersection or actual relationships among them to base this trust on.
Large-scale online voting is eschewed as every federation, no matter how large,  is governed by an assembly that engages in small-scale democracy, where votes should be taken by a show of hands for transparency, allowing people to see how their representatives have voted.

\begin{figure}[ht]
  \begin{center}
   \includegraphics[width=12cm]{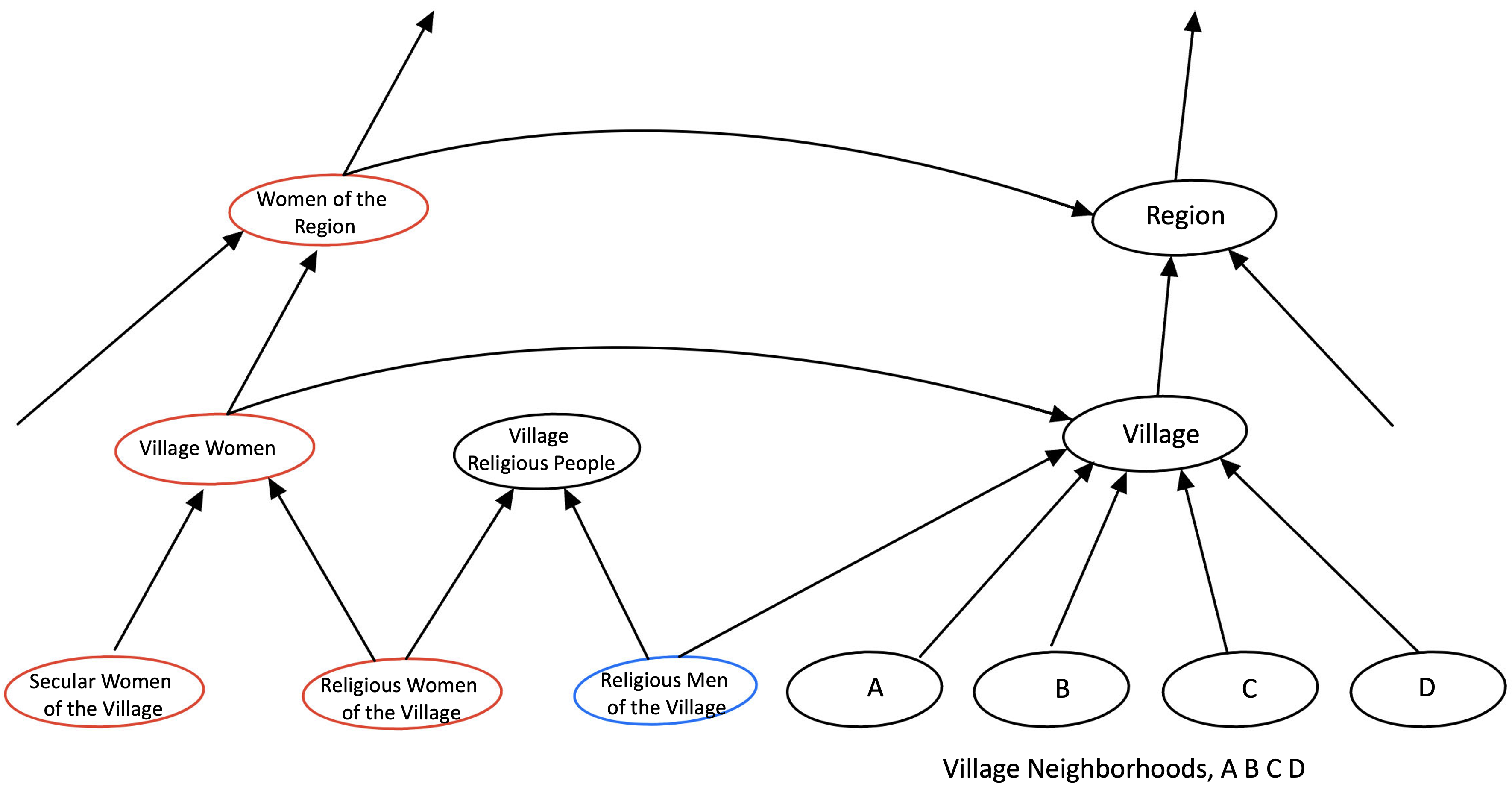}
  \end{center}
\caption{\textbf{Grassroots Federation.} An imaginary grassroots federation of the people of a village, which has four neighbourhoods (A-D). People have assembled into various communities of interest (secular and religious woman, religious man), of which the four neighbourhoods, the women and the religious men were admitted as child federations/communities to the village federation.  The village is a member of the regional federation.  The federation of the women of the village is a member of both the village federation and the regional women's federation, which in turn is both a member of the regional federation as well as the higher (national?) women's federation.}
\label{figure:federated-assemblies} 
\end{figure}
The underlying structure of the federation can represent both geographic communities (e.g., residents of a neighbourhood) and communities of interest (e.g., association for the protection of animal rights). See Figure \ref{figure:federated-assemblies}.

A person is expected to be a member of a single geographic community (e.g. a neighbourhood), which may federate hierarchically.  However, a person may be a member of several communities of interests (e.g. women association, animal rights association, religious association).  A federation may include both geographic child communities/federations (e.g.neighbourhoods of a village), and communities/federations of interest (e.g. the village chapter of the women association).  The challenge of selection by sortition of an assembly of a federation is to ensure proportional representation for all child assemblies (which may be overlapping in case of communities of interest) while ensuring equal representation to each person (who may be a member of a geographic community as well as several communities of interest), regardless of the number and types of communities the person is a member of.

\section{Conclusions}\label{section:conclusions}
We have presented a grassroots architecture with the potential to supplant global digital platforms with a global digital democracy.
The challenge to realise it as a common good for the benefit of all remains open.

\subsection*{Acknowledgements}
The project described here was initiated by the Digital Democracy group at Weizmann, notably with Nimrod Talmon, Gal Shahaf, and Ouri Poupko, and later expanded to collaborations with many colleagues,  all of whom are co-authors of the papers referenced herein.
Ehud Shapiro is the Incumbent of The Harry Weinrebe Professorial Chair of Computer Science and Biology at the Weizmann Institute and a Visiting Professor at the London School of Economics.

\bibliographystyle{splncs04}
\bibliography{bib}

\begin{thebibliography}{10}
\providecommand{\url}[1]{\texttt{#1}}
\providecommand{\urlprefix}{URL }
\providecommand{\doi}[1]{https://doi.org/#1}

\bibitem{abramowitz2021democratic}
Abramowitz, B., Elkind, E., Grossi, D., Shapiro, E., Talmon, N.: Democratic forking: Choosing sides with social choice. In: International Conference on Algorithmic Decision Theory. pp. 341--356. Springer (2021)

\bibitem{abramowitz2021beginning}
Abramowitz, B., Shapiro, E., Talmon, N.: In the beginning there were $n$ agents: Founding and amending a constitution. In: Proceedings of ADT '21. pp. 119--131 (2021)

\bibitem{allen2016path}
Allen, C.: {The Path to Self-Sovereign Identity} (2016, http://wwwlifewithalacritycom/2016/04/the-path-to-self-soverereign-identityhtml), \url{http://www.lifewithalacrity.com/2016/04/the-path-to-self-soverereign-identity.html}, [accessed 2022]

\bibitem{almeida2024blocklace}
Almeida, P.S., Shapiro, E.: The blocklace: A universal, byzantine fault-tolerant, conflict-free replicated data type. arXiv preprint arXiv:X.X  (2024)

\bibitem{alvisi2013sok}
Alvisi, L., Clement, A., Epasto, A., Lattanzi, S., Panconesi, A.: Sok: The evolution of sybil defense via social networks. In: Proceedings of the 2013 {IEEE} symposium on security and privacy. pp. 382--396. IEEE (2013)

\bibitem{auvolat2020money}
Auvolat, A., Frey, D., Raynal, M., Ta{\"\i}ani, F.: Money transfer made simple: a specification, a generic algorithm, and its proof. arXiv preprint arXiv:2006.12276  (2020)

\bibitem{babel2023mysticeti}
Babel, K., Chursin, A., Danezis, G., Kokoris-Kogias, L., Sonnino, A.: Mysticeti: Low-latency dag consensus with fast commit path. arXiv preprint arXiv:2310.14821  (2023)

\bibitem{borge2017proof}
Borge, M., Kokoris-Kogias, E., Jovanovic, P., Gasser, L., Gailly, N., Ford, B.: Proof-of-personhood: Redemocratizing permissionless cryptocurrencies. In: 2017 IEEE European Symposium on Security and Privacy Workshops (EuroS\&PW). pp. 23--26. IEEE (2017)

\bibitem{bracha1987asynchronous}
Bracha, G.: Asynchronous {B}yzantine agreement protocols. Information and Computation  \textbf{75}(2),  130--143 (1987)

\bibitem{bulteau2021aggregation}
Bulteau, L., Shahaf, G., Shapiro, E., Talmon, N.: Aggregation over metric spaces: Proposing and voting in elections, budgeting, and legislation. Journal of Artificial Intelligence Research  \textbf{70},  1413--1439 (2021)

\bibitem{buterin2014next}
Buterin, V., et~al.: A next-generation smart contract and decentralized application platform. white paper  \textbf{3}(37), ~2--1 (2014)

\bibitem{cardelli2020digital}
Cardelli, L., Orgad, L., Shahaf, G., Shapiro, E., Talmon, N.: Digital social contracts: A foundation for an egalitarian and just digital society. In: CEUR Proceedings of the First International Forum on Digital and Democracy. vol.~2781, pp. 51--60. CEUR-WS (2020)

\bibitem{collins2020online}
Collins, D., Guerraoui, R., Komatovic, J., Kuznetsov, P., Monti, M., Pavlovic, M., Pignolet, Y.A., Seredinschi, D.A., Tonkikh, A., Xygkis, A.: Online payments by merely broadcasting messages. In: 2020 50th Annual IEEE/IFIP International Conference on Dependable Systems and Networks (DSN). pp. 26--38. IEEE (2020)

\bibitem{de2021smart}
De~Filippi, P., Wray, C., Sileno, G.: Smart contracts. Internet Policy Review  \textbf{10}(2) (2021)

\bibitem{douceur2002sybil}
Douceur, J.R.: The sybil attack. In: Proceedings of the international workshop on peer-to-peer systems. pp. 251--260 (2002)

\bibitem{elkind2022complexity}
Elkind, E., Ghosh, A., Goldberg, P.: Complexity of deliberative coalition formation. arXiv preprint arXiv:2202.12594  (2022)

\bibitem{elkind2021united}
Elkind, E., Grossi, D., Shapiro, E., Talmon, N.: United for change: Deliberative coalition formation to change the status quo. In: Proceedings of AAAI '21. vol.~35, pp. 5339--5346 (2021)

\bibitem{ethereum:dao}
Ethereum: {Decentralized autonomous organizations (DAOs) | ethereum.org} (2021, https://ethereumorg/en/dao), \url{{https://ethereum.org/en/dao/}}, [accessed 09-December-2021]

\bibitem{giridharan2022bullshark}
Giridharan, N., Kokoris-Kogias, L., Sonnino, A., Spiegelman, A.: Bullshark: {DAG} {BFT} protocols made practical. arXiv preprint arXiv:2201.05677  (2022)

\bibitem{guerraoui2019consensus}
Guerraoui, R., Kuznetsov, P., Monti, M., Pavlovi{\v{c}}, M., Seredinschi, D.A.: The consensus number of a cryptocurrency. In: Proceedings of the 2019 ACM Symposium on Principles of Distributed Computing. pp. 307--316 (2019)

\bibitem{halpern2024federated}
Halpern, D., Procaccia, A.D., Shapiro, E., Talmon, N.: Federated assemblies. arXiv preprint arXiv:2405.19129  (2024)

\bibitem{halpern2024grassroots}
Halpern, D., Proccacia, A., Shapiro, E., Talmon, N.: Grassroots federated assemblies. In preparation  (2024)

\bibitem{DBLP:conf/sicherheit/JacobBH22}
Jacob, F., Bayreuther, S., Hartenstein, H.: On crdts in byzantine environments. In: Wressnegger, C., Reinhardt, D., Barber, T., Witt, B.C., Arp, D., Mann, Z.{\'{A}}. (eds.) Sicherheit, Schutz und Zuverl{\"{a}}ssigkeit: Konferenzband der 11. Jahrestagung des Fachbereichs Sicherheit der Gesellschaft f{\"{u}}r Informatik e.V. (GI), Sicherheit 2022, Karlsruhe, Germany, April 5-8, 2022. {LNI}, vol. {P-323}, pp. 113--126. Gesellschaft f{\"{u}}r Informatik, Bonn (2022). \doi{10.18420/SICHERHEIT2022\_07}, \url{https://doi.org/10.18420/sicherheit2022\_07}

\bibitem{keidar2021need}
Keidar, I., Kokoris-Kogias, E., Naor, O., Spiegelman, A.: All you need is dag. In: Proceedings of the 2021 ACM Symposium on Principles of Distributed Computing. pp. 165--175 (2021)

\bibitem{keidar2023cordial}
Keidar, I., Naor, O., Shapiro, E.: Cordial miners: A family of simple and efficient consensus protocols for every eventuality. In: 37th International Symposium on Distributed Computing (DISC 2023). LIPICS (2023)

\bibitem{DBLP:conf/eurosys/Kleppmann22}
Kleppmann, M.: Making crdts byzantine fault tolerant. In: Szekeres, A., Sivaramakrishnan, K.C. (eds.) PaPoC@EuroSys 2022: Proceedings of the 9th Workshop on Principles and Practice of Consistency for Distributed Data, Rennes, France, April 5 - 8, 2022. pp. 8--15. {ACM} (2022). \doi{10.1145/3517209.3524042}, \url{https://doi.org/10.1145/3517209.3524042}

\bibitem{DBLP:journals/corr/abs-2012-00472}
Kleppmann, M., Howard, H.: Byzantine eventual consistency and the fundamental limits of peer-to-peer databases. CoRR  \textbf{abs/2012.00472} (2020), \url{https://arxiv.org/abs/2012.00472}

\bibitem{lewenberg2015inclusive}
Lewenberg, Y., Sompolinsky, Y., Zohar, A.: Inclusive block chain protocols. In: Financial Cryptography and Data Security: 19th International Conference, FC 2015, San Juan, Puerto Rico, January 26-30, 2015, Revised Selected Papers 19. pp. 528--547. Springer (2015)

\bibitem{RN47}
Lewis, A.: A gentle introduction to self-sovereign identity  (2017), \url{{https://bitsonblocks.net/2017/05/17/a-gentle-introduction-to-self-sovereign-identity/}}

\bibitem{lewispye2023flash}
Lewis-Pye, A., Naor, O., Shapiro, E.: Flash: An asynchronous payment system with good-case linear communication complexity. arXiv preprint arXiv:2305.03567  (2023)

\bibitem{lewis2023grassroots}
Lewis-Pye, A., Naor, O., Shapiro, E.: Grassroots flash: A payment system for grassroots cryptocurrencies. arXiv preprint arXiv:2309.13191  (2023)

\bibitem{lewis2024goodfellas}
Lewis-Pye, A., Shapiro, E.: Goodfellas consensus: Constant latency with linear complexity and throughput. In preparation  (2024)

\bibitem{lewis2024laidback}
Lewis-Pye, A., Shapiro, E.: Ladiback consensus: No leaders, no rounds, no conflicts, no overhead. In preparation  (2024)

\bibitem{lichtenstein1988concurrent}
Lichtenstein, Y., Shapiro, E.: Concurrent algorithmic debugging. ACM SIGPLAN Notices  \textbf{24}(1),  248--260 (1988)

\bibitem{may1952set}
May, K.O.: A set of independent necessary and sufficient conditions for simple majority decision. Econometrica: Journal of the Econometric Society pp. 680--684 (1952)

\bibitem{meir2020sybil}
Meir, R., Shahaf, G., Shapiro, E., Talmon, N.: Sybil-resilient social choice with partial participation. arXiv preprint arXiv:2001.05271  (2020)

\bibitem{meir2024safe}
Meir, R., Shahaf, G., Shapiro, E., Talmon, N.: Safe voting: Resilience to abstention and sybils. arXiv preprint arXiv:2001.05271  (2024)

\bibitem{bitcoin}
Nakamoto, S.: Bitcoin: A peer-to-peer electronic cash system (2008), \url{https://bitcoin.org/bitcoin.pdf}

\bibitem{park2021going}
Park, S., Specter, M., Narula, N., Rivest, R.L.: Going from bad to worse: from internet voting to blockchain voting. Journal of Cybersecurity  \textbf{7}(1),  tyaa025 (2021)

\bibitem{proudhon2004general}
Proudhon, P.J., Robinson, J.B.: General idea of the revolution in the nineteenth century. Courier Corporation (2004)

\bibitem{reijers2018now}
Reijers, W., Wuisman, I., Mannan, M., De~Filippi, P., Wray, C., Rae-Looi, V., V{\'e}lez, A.C., Orgad, L.: Now the code runs itself: On-chain and off-chain governance of blockchain technologies. Topoi pp. 1--11 (2018)

\bibitem{safra1988meta}
Safra, S., Shapiro, E.: Meta interpreters for real. In: Concurrent Prolog: Collected Papers, pp. 166--179. MIT Press (1988)

\bibitem{seuken2014sybil}
Seuken, S., Parkes, D.C.: Sybil-proof accounting mechanisms with transitive trust. In: Proceedings of the 2014 international conference on Autonomous agents and multi-agent systems. pp. 205--212. International Foundation for Autonomous Agents and Multiagent Systems (2014)

\bibitem{shahaf2019sybil}
Shahaf, G., Shapiro, E., Talmon, N.: Sybil-resilient reality-aware social choice. In: Proceedings of the 28th International Joint Conference on Artificial Intelligence. pp. 572--579 (2019)

\bibitem{shapiro1987concurrent}
Shapiro, E.: Concurrent prolog: Collected papers (1987)

\bibitem{shapiro1989family}
Shapiro, E.: The family of concurrent logic programming languages. ACM Computing Surveys (CSUR)  \textbf{21}(3),  413--510 (1989)

\bibitem{shapiro2021multiagent}
Shapiro, E.: Multiagent transition systems: Protocol-stack mathematics for distributed computing. arXiv preprint arXiv:2112.13650  (2021)

\bibitem{shapiro2023grassrootsBA}
Shapiro, E.: Grassroots distributed systems: Concept, examples, implementation and applications (brief announcement). In: 37th International Symposium on Distributed Computing (DISC 2023). LIPICS (2023)

\bibitem{shapiro2023gsn}
Shapiro, E.: Grassroots social networking: Serverless, permissionless protocols for twitter/linkedin/whatsapp. In: OASIS ’23. Association for Computing Machinery (2023). \doi{10.1145/3599696.3612898}

\bibitem{shapiro2023grassroots}
Shapiro, E.: Grassroots systems: Concept, examples, implementation and applications. arXiv preprint arXiv:2301.04391  (2023)

\bibitem{shapiro2024gc}
Shapiro, E.: Grassroots currencies: Foundations for grassroots digital economies. arXiv preprint arXiv:2202.05619  (2024)

\bibitem{shapiro2024gpl}
Shapiro, E.: Grassroots programming languages. In preparation  (2024)

\bibitem{shapiro2018incorporating}
Shapiro, E., Talmon, N.: Incorporating reality into social choice. In: Proceedings of the 17th International Conference on Autonomous Agents and MultiAgent Systems. pp. 1188--1192 (2018)

\bibitem{shapiro2022foundations}
Shapiro, E., Talmon, N.: Foundations for grassroots democratic metaverse. arXiv preprint arXiv:2203.04090  (2022)

\bibitem{DBLP:conf/sss/ShapiroPBZ11}
Shapiro, M., Pregui{\c{c}}a, N.M., Baquero, C., Zawirski, M.: Conflict-free replicated data types. In: D{\'{e}}fago, X., Petit, F., Villain, V. (eds.) Stabilization, Safety, and Security of Distributed Systems - 13th International Symposium, {SSS} 2011, Grenoble, France, October 10-12, 2011. Proceedings. Lecture Notes in Computer Science, vol.~6976, pp. 386--400. Springer (2011). \doi{10.1007/978-3-642-24550-3\_29}

\bibitem{siddarth2020watches}
Siddarth, D., Ivliev, S., Siri, S., Berman, P.: Who watches the watchmen? a review of subjective approaches for sybil-resistance in proof of personhood protocols. Frontiers in Blockchain  \textbf{3}, ~46 (2020)

\bibitem{sompolinsky2016spectre}
Sompolinsky, Y., Lewenberg, Y., Zohar, A.: Spectre: A fast and scalable cryptocurrency protocol. Cryptology ePrint Archive  (2016)

\bibitem{sompolinsky2022dag}
Sompolinsky, Y., Sutton, M.: The dag knight protocol: A parameterless generalization of nakamoto consensus. Cryptology ePrint Archive  (2022)

\bibitem{sompolinsky2021phantom}
Sompolinsky, Y., Wyborski, S., Zohar, A.: Phantom ghostdag: a scalable generalization of nakamoto consensus: September 2, 2021. In: Proceedings of the 3rd ACM Conference on Advances in Financial Technologies. pp. 57--70 (2021)

\bibitem{sterling1994art}
Sterling, L., Shapiro, E.: The art of Prolog: advanced programming techniques. MIT press (1994)

\bibitem{tromer2018benaloh}
Tromer, E.: {Report on criticism of blockchain democracy by Josh Benaloh} (2018, https://githubcom/ZcashFoundation/GrantProposals-2018Q2/issues/22\#issuecomment-392825397)

\bibitem{RN592}
Wang, F., De~Filippi, P.: Self-sovereign identity in a globalized world: Credentials-based identity systems as a driver for economic inclusion. Frontiers in Blockchain  \textbf{2}, ~28 (2020)

\bibitem{wang2018overview}
Wang, S., Yuan, Y., Wang, X., Li, J., Qin, R., Wang, F.Y.: An overview of smart contract: architecture, applications, and future trends. In: 2018 IEEE Intelligent Vehicles Symposium (IV). pp. 108--113. IEEE (2018)

\bibitem{webb2018fork}
Webb, N.: A fork in the blockchain: Income tax and the bitcoin/bitcoin cash hard fork. North Carolina Journal of Law \& Technology  \textbf{19}(4), ~283 (2018)

\bibitem{zuboff2019age}
Zuboff, S.: The age of surveillance capitalism: The fight for a human future at the new frontier of power. Public Affairs (2019)

\end{thebibliography}

\end{document}